\documentclass[aps,showpacs,preprint,footinbib,preprintnumbers]{revtex4}
\usepackage{amssymb}
\usepackage{amsmath}
\usepackage{amsfonts}
\usepackage{bm}
\usepackage{here}
\usepackage{graphicx}

\newcommand{\be}{\begin{equation}}
\newcommand{\ee}{\end{equation}}
\newcommand{\ba}{\begin{eqnarray}}
\newcommand{\ea}{\end{eqnarray}}
\newcommand{\nn}{\nonumber}

\begin{document}


\title{ Study of $X(5568)$ in a unitary coupled-channel approximation of $B \bar{K}$ and $B_s \pi$}

\author{Bao-Xi Sun$^{1,2}$, Fang-Yong Dong$^{1}$ and Jing-Long Pang$^{2}$}
\affiliation{$^{1}$College of Applied Sciences, Beijing University
of Technology, Beijing 100124, China}
\affiliation{$^{2}$Department of Physics, Peking University,
Beijing 100871, China}


\begin{abstract}
The potential of the $B$ meson and the pseudoscalar meson is constructed up to the next-to-leading order Lagrangian, and then the $B \bar{K}$ and $B_s \pi$ interaction  is studied in the unitary coupled-channel approximation, and a resonant state with a mass about $5568MeV$ and $J^P=0^+$ is generated dynamically, which can be associated with the $X(5568)$ state announced by D0 Collaboration recently. The mass and the decay width of this resonant state depend on the regularization scale in the dimensional regularization scheme, or the maximum momentum in the momentum cutoff regularization scheme. The scattering amplitude of the vector $B$ meson and the pseudoscalar meson is calculated, and an axial-vector state with a mass near $5620MeV$ and $J^P=1^+$ is produced.
Moreover, their partners in the charm sector are also discussed.
%
\end{abstract}


\maketitle


\section{Introduction}

The D0 Collaboration announce the discovery of a new state,
$X(5568)$, as a narrow peak in the $B^0_s \pi^{\pm}$ invariant mass
distribution with significance of $5.1\sigma$ based on $10.4fb^{-1}$
of $p\bar{p}$ collision at $\sqrt{s}=1.96TeV$\cite{D0}. The $X(5568)$
has a mass of $M=5567.8 \pm 2.9^{+0.9}_{-1.9} MeV$ and a decay width of
$\Gamma= 21.9 \pm 6.4 ^{+5.0}_{-2.5} MeV$.
This discovery implies $X(5568)$ might be composed of two
quarks and two antiquarks, of four different flavors: $b$, $s$, $u$, $d$,
and thus it might be a candidate of the exotic tetraquark states.
Along this clue, many theorists try to explain its properties at
the quark level, or treat it as a $molecule$ state of a $B$ meson
and a pseudoscalar meson
\cite{Agaev:2016mjb,Wang:2016tsi,Wang:2016mee,Chen:2016mqt,Liu:2016xly,Agaev:2016ijz,Liu:2016ogz,Agaev:2016lkl,Dias:2016dme,
Wang:2016wkj,Agaev:2016urs,He:2016yhd,Stancu:2016sfd,Burns:2016gvy,Tang:2016pcf,Guo:2016nhb,Esposito:2016itg,Albaladejo:2016eps,
 Lu:2016kxm, Goerke:2016hxf, Agaev:2016ifn}.
However, some theoretical works do not support the existence of the $X(5568)$ state\cite{Zanetti:2016wjn,Jin:2016cpv,Lu:2016zhe,Albuquerque:2016nlw,Chen:2016npt,
Kang:2016zmv,Lang:2016jpk,Chen:2016ypj}.
Recently, a negative result has been reported by the LHCb
Collaboration\cite{LHCb:2016ppf}. They claim that no significant
excess has been found, based on the data sample recorded with the LHCb
detector corresponding to $3fb^{-1}$ of $pp$ collision data at
$\sqrt{s}=7\sim8 TeV$.
In addition, the CMS Collaboration search for
the resonance-like structures in the $B_s^0 \pi^{\pm}$ invariant
mass spectrum using an integrated luminosity of $19.7fb^{-1}$ of
proton-proton collisions at $\sqrt{s}=8TeV$, and they declare no
hint for the $X(5568)$ particle is shown\cite{CMS}.

In the present work, the interaction of the $B$ meson and the
pseudoscalar meson is studied in the chiral perturbation theory. We
calculate the interaction potentials up to next-to-leading order
correction, and then solve the Bethe-Salpeter equation in the
unitary coupled-channel approximation. A resonant state with a mass
about $5568MeV$ is generated dynamically in the $B \bar{K}$ and $B_s
\pi$ channels, which is assumed to be associated with the $X(5568)$
particle. Thus we think that the $X(5568)$ state has a quantum number
$J^P=0^+$.

This article is organized as follows. In Section~\ref{sect:chiral},
the effective chiral Lagrangian is given. In section~\ref{sect:Bpi},
the potential of the $B$ meson and the pseudoscalar meson up to the
next-to-leading order correction is constructed according to the
Lagrangian in Section~\ref{sect:chiral}. In Section~\ref{sect:BS},
the Bethe-Salpeter equation is discussed in the unitary
coupled-channel approximation. The calculation results on the $B_s
\pi$ and $B \bar{K}$ channels are summarized in
Section~\ref{sect:result}.
 In Section~\ref{sect:Bstar}, the coupled channels $B^*_s \pi$ and $B^* \bar{K}$ are studied in the unitary coupled-channel approximation, and a resonant state with a mass near $5620MeV$ and $J^P=1^+$ is generated dynamically.
In Section~\ref{sect:charm}, the corresponding charm sectors are analyzed in the unitary coupled-channel approximation.
Finally, a summary is given in Section~\ref{sect:summary}.

\section{Chiral Lagrangian up to the next-to-leading order}
\label{sect:chiral}

For the $B$ meson, the triplets take forms of
$P=(B^-,\bar{B}^0,\bar{B}_s^0)$ and $P^\dagger=(B^+,B^0,B_s^0)$, and
for the vector $B^*$ meson, we have
$P_\mu^*=({B^*}^-,\bar{B^*}^0,\bar{B^*}_s^0)_\mu$ and ${P^*}_\mu^\dagger=({B^*}^+,{B^*}^0,{B_s^*}^0)_\mu$.
 The leading order chiral Lagrangian describing the interactions of
the $B$ meson with the pseudoscalar meson can be
written as:
\begin{eqnarray}
\label{eq:lolag}
\mathcal{L}^{(1)}&=&\langle\mathcal{D}_\mu
P\mathcal{D}^\mu P^\dagger\rangle -m_{P}^2\langle PP^\dagger\rangle
-\langle\mathcal{D}_\mu P^{* \nu} \mathcal{D}^\mu P^{*\dagger}_\nu
\rangle + m_{P^*}^2\langle P^{* \nu} P_\nu^{* \dagger}\rangle ,
\end{eqnarray}
where $m_{P}$ and $m_{P^*}$ are the masses of the $B$ and vector $B^*$ meson, respectively, and $\langle
...\rangle$ denotes the trace in the $SU(3)$ flavor space. The
chiral covariant derivative is defined as \be \mathcal{D}_\mu P_a =
\partial_\mu P_a - \Gamma^{ba}_\mu P_b, ~~~~ \mathcal{D}^\mu
P^\dagger_a =
\partial^\mu P^\dagger_a - \Gamma_{ab}^\mu P^\dagger_b,
\ee where the vector current
$\Gamma_\mu=\frac{1}{2}(\xi^\dagger\partial_\mu \xi+\xi\partial_\mu
\xi^\dagger)$ and $\xi^2=\exp(i\Phi/f_0)$ with $f_0$ being
 the decay constant of the pseudoscalar meson and $\Phi$
the octet of pseudoscalar mesons:
\begin{equation}
\Phi=\sqrt{2}\left(
\begin{array}{ccc}
\frac{\pi^0}{\sqrt{2}}+\frac{\eta}{\sqrt{6}} & \pi^+ & K^+\\
\pi^- & -\frac{\pi^0}{\sqrt{2}}+\frac{\eta}{\sqrt{6}} & K^0\\
K^- & \bar{K}^0 & -\frac{2}{\sqrt{6}}\eta
\end{array}
\right).
\end{equation}

Similarly, the covariant next-to-leading order(NLO) terms of the
effective Lagrangian are constructed:
\begin{eqnarray}
\label{eq:nlolag}
\mathcal{L}^{(2)}&=&-2[c_0\langle P P^\dagger\rangle\langle \chi_+\rangle-c_1\langle P \chi_+ P^\dagger\rangle-c_2\langle P P^\dagger\rangle\langle u^\mu u_\mu\rangle-c_3\langle P u^\mu u_\mu P^\dagger\rangle\nonumber\\&&+\frac{c_4}{m_P^2}\langle \mathcal D_\mu P\mathcal D_\nu P^\dagger\rangle\langle\{u^\mu,u^\nu\}\rangle+\frac{c_5}{m_P^2}\langle \mathcal D_\mu P\{u^\mu,u^\nu\} \mathcal D_\nu P^\dagger\rangle+\frac{c_6}{m_P^2}\langle \mathcal D_\mu P[u^\mu,u^\nu]\mathcal D_\nu P^\dagger\rangle]\nonumber\\
&&+2[\tilde{c}_0\langle P^*_\mu P^{* \mu \dagger}\rangle\langle
\chi_+\rangle-\tilde{c}_1\langle P^*_\mu \chi_+ P^{* \mu \dagger}
\rangle-\tilde{c}_2\langle P^*_\nu P^{* \nu \dagger} \rangle\langle
u^\mu u_\mu\rangle-\tilde{c}_3\langle P^*_\nu u^\mu u_\mu P^{* \nu
\dagger} \rangle \nonumber\\
&&+\frac{\tilde{c}_4}{m_{P^*}^2}\langle \mathcal D_\mu P^*_\alpha
\mathcal D_\nu P^{* \alpha \dagger}
\rangle\langle\{u^\mu,u^\nu\}\rangle+\frac{\tilde{c}_5}{m_{P^*}^2}\langle
\mathcal D_\mu P^*_\alpha \{u^\mu,u^\nu\} \mathcal D_\nu P^{* \alpha
\dagger} \rangle \nn \\
&&+\frac{\tilde{c}_6}{m_{P^*}^2}\langle \mathcal D_\mu P^*_\alpha
[u^\mu,u^\nu]\mathcal D_\nu P^{* \alpha \dagger} \rangle],
\end{eqnarray}
where $u_\mu=i(\xi^\dagger\partial_\mu \xi-\xi\partial_\mu
\xi^\dagger)$ being the axial current and $\chi_+=\xi^\dagger
\mathcal{M} \xi^\dagger + \xi\mathcal{M}\xi$ with $\mathcal{M}=
\mathrm{diag}(m_\pi^2,m_\pi^2,2 m_K^2-m_\pi^2)$\cite{Guo:2008gp,Altenbuchinger:2013vwa}.

\section{The potentials for $B \phi \rightarrow B \phi$}

\label{sect:Bpi}

For the process of $B_1(p_1)+\phi_1(k_1)\rightarrow B_2(p_2)+
\phi_2(k_2)$, the leading order potential can be written as \be
V_{LO}= \frac{1}{8f_0^2}C_{LO} (s-u)  \ee with the Mandelstam
variables $s=(k_1+p_1)^2=(k_2+p_2)^2$ and
$u=(k_2-p_1)^2=(k_1-p_2)^2$. The coefficient $C_{LO}$ for the
different channels are listed in Table~\ref{table:C-LO}.

The next-to-leading order potential between the $B$ meson and the
pseudoscalar meson takes the following form:
\ba \label{eq:V_NLO} V_{NLO} &=& \frac{1}{f_0^2} c_0 C_0 -
\frac{1}{4f_0^2} c_1 C_1
\nn \\
&+&\frac{2}{f_0^2}c_2 C_2 k_1 \cdot k_2
+\frac{1}{f_0^2}c_3 C_3 k_1 \cdot k_2 \nn \\
&-&\frac{4}{m_P^2 f_0^2} c_4 * \left( C_{41}p_1 \cdot k_1 p_2 \cdot
k_2 + C_{42} p_1 \cdot k_2 p_2 \cdot k_1 \right) \nn \\
&-&\frac{2}{m_P^2 f_0^2} c_5 * \left( C_{51}p_1 \cdot k_1 p_2 \cdot
k_2 + C_{52} p_1 \cdot k_2 p_2 \cdot k_1 \right).
 \ea

We shall discuss the amplitudes using the isospin formalism, and the
state with isospin $I=1$ can be written as
\be |B \bar{K}, I=1 \rangle = -\sqrt{\frac{1}{2}} B^+ K^-
+\sqrt{\frac{1}{2}}B^0 \bar{K}^0,
\ee where the phase convention $|K^- \rangle = -|\frac{1}{2},
-\frac{1}{2} \rangle$ for the isospin state has been used.

\begin{table}[hbt]
\begin{center}
\begin{tabular}{c|ccccccccc}
\hline
 & $C_{LO}$ & $C_0$ & $C_1$ & $C_2$ & $C_3$ & $C_{41}$ & $C_{42}$ & $C_{51}$&$C_{52}$  \\
\hline $B \bar{K}\rightarrow B \bar{K}$ & $0$ & $4 m_K^2$ & $0$ &
$-2$ & $0$ &
$-1$ & $-1$ & $0$&$0$  \\
$B \bar{K}\rightarrow B^0_s \pi^0$ & $-2$ & $0$
& $4(m_K^2+m_\pi^2)$ & $0$ & $2$ & $0$ & $0$ & $0$&$2$  \\
$B^0_s \pi^0\rightarrow B^0_s \pi^0$ & $0$ & $4 m_\pi^2$ & $0$ &
$-2$ & $0$ &
$-1$ & $-1$ & $0$&$0$  \\
\hline
\end{tabular}
\caption{Coefficients for the channels with $B=1$, $S=-1$ and $I=1$.}
\label{table:C-LO}
\end{center}
\end{table}

In the heavy-meson chiral perturbation theory, the leading order
interaction between the B meson and the pseudoscalar meson can be
written as
\be
\label{eq:LO}
 V_{LO}= \frac{m_B}{4f^2_0} C_{LO} \left( E+E^\prime \right)
\ee
with $E$ and $E^\prime$ the energies of the initial an final
pseudoscalar mesons, respectively.

The next-to-leading order interaction can be deduced from
Eq.~(\ref{eq:V_NLO}) in the heavy-meson approximation
\be
V_{NLO}=-\frac{2}{f_0^2}c_0 C_0 + \frac{1}{2f_0^2} c_1 C_1
-\frac{4}{f_0^2} c_{24} C_2 E E^\prime
-\frac{2}{f_0^2} c_{35} C_3 E E^\prime,
\ee
where $c_{24}=c_2-2c_4$ and $c_{35}=c_3-2c_5$.

\begin{figure}[htb]
\begin{center}
\includegraphics[width=0.8\textwidth]{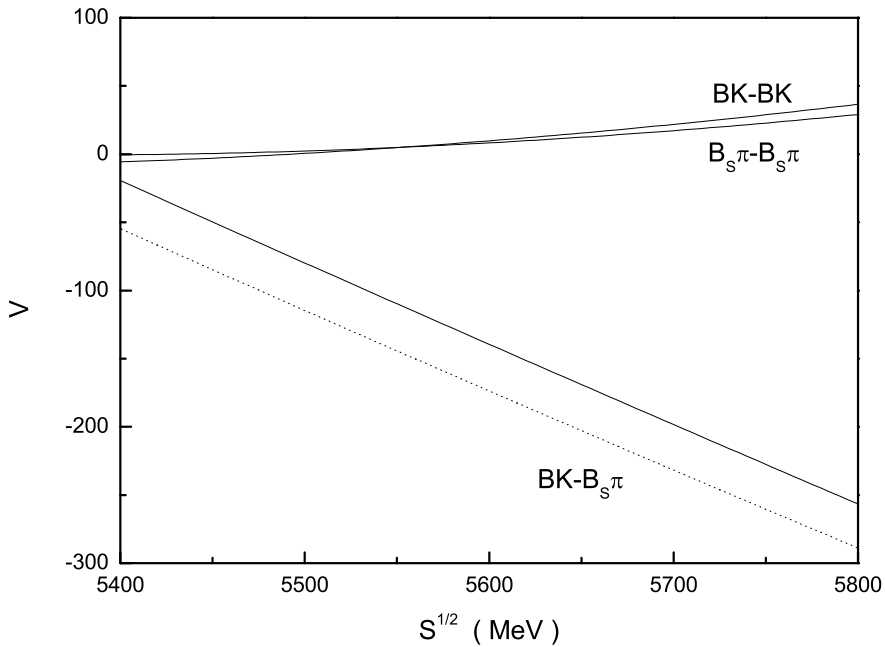}
\end{center}
\caption{The potential of the $B$ meson and the pseudoscalar meson
in the different channels. The solid lines denote the potentials in
the leading order plus next-to-leading order approximation, and the
dotted line stands for the leading order potential in the $B \bar{K}
\rightarrow B_s \pi$ channel.
} \label{fig:V_0629}
\end{figure}

Up to the next-to-leading order Lagrangian, six low energy constants
should be fixed. These low energy constants have been determined
according to the Lattice QCD data in Ref.~\cite{Liu:2012zya} when
the interaction between the pseudoscalar meson and the $D$ meson is
studied in a coupled-channel approximation of the chiral unitary
theory, and the $D^*_{s0}(2317)$ is produced in the $DK$
channel\cite{Altenbuchinger:2013vwa}. The values of them are
depicted in Table~\ref{tab:LEC}. Since the term multiplying $c_6$
is in fact of $O(p^3)$ order, and it has negligible effects on the
dynamically generation of resonant states as long as $c_6$ is in a
natural size, we set $c_6$ to be zero just as in
Ref.~\cite{Guo:2008gp,Altenbuchinger:2013vwa}.

\begin{table}[hbt]
\begin{center}
\begin{tabular}{c|cccc}
\hline
 & $c_0$ & $c_1$ & $c_{24}$ & $c_{35}$  \\
\hline
HQS UChPT        & $0.015$    & $-0.214$ & $-0.068$ & $-0.011$    \\
\hline
\end{tabular}
\caption{Low energy constants for the pseudoscalar meson and $D$
meson interaction.
} \label{tab:LEC}
\end{center}
\end{table}

The low energy constants $c_0$,...,$c_6$ in the interaction of the $B$
meson and the pseudoscalar meson are related to those values in the
case of the $D$ meson and the pseudoscalar meson interaction. Up to
corrections in $1/m_B(m_D)$,
\be c_{i,B}/m_B=c_{i,D}/m_D, \ee
where $m_B=5331.9MeV$ and $m_D=1972.1MeV$ are the average masses of
$B$ and $D$ mesons, respectively.

\section{Unitary coupled-channel approximation of Bethe-Salpeter equation}
\label{sect:BS}

The amplitude can be obtained by solving
Bethe-Salpeter equation in the S-wave approximation,
   \begin{equation}
T(\sqrt{s}) = [1 - V_S(\sqrt{s}) \, G(s)]^{-1}\, V_S(\sqrt{s}),
\label{eq:Bethe}
\end{equation}
which is a function of the total energy $\sqrt{s}$ in the center of
mass system. In Eq.~(\ref{eq:Bethe}), $G(s)$ is the propagator of a
pseudoscalar meson and a $B$ meson, and it can be calculated
explicitly in a dimensional regularization scheme\cite{ollerulf}:

\begin{eqnarray}
G_{l}(s) &=& i  \int \frac{d^4 q}{(2 \pi)^4} \, \frac{1}{(P-q)^2 -
M_l^2 + i \epsilon} \, \frac{1}{q^2 - m^2_l + i \epsilon} \nonumber
\\ &=& \frac{1}{16 \pi^2} \left\{ a_l
+ \ln \frac{M_l m_l}{\mu^2} + \frac{m_l^2-M_l^2}{2s} \ln
\frac{m_l^2}{M_l^2}  \right. \nonumber \\
& &
%
+ \frac{\bar{q}_l}{\sqrt{s}} \left[
\ln(s-(M_l^2-m_l^2)+2\bar{q}_l\sqrt{s})+
\ln(s+(M_l^2-m_l^2)+2\bar{q}_l\sqrt{s}) \right.   \\
& & \left. \phantom{\frac{2 M}{16 \pi^2} +
\frac{\bar{q}_l}{\sqrt{s}}} \left. \hspace*{-0.3cm}-
\ln(-s+(M_l^2-m_l^2)+2\bar{q}_l\sqrt{s})-
\ln(-s-(M_l^2-m_l^2)+2\bar{q}_l\sqrt{s}) \right] \right\} \ , \nonumber
\label{eq:gpropdr}
\end{eqnarray}
with $\mu$ the regularization scale, and $a_l$ the subtraction constant.

In eq.~(\ref{eq:gpropdr}), $\bar{q}_l$ denotes the three-momentum of
mesons in the center of mass system and is given by
\begin{equation}
\bar{q}_l=\frac{\sqrt{s-(M_l+m_l)^2}\sqrt{s-(M_l-m_l)^2}}{2\sqrt{s}},
\end{equation}
where $M_l$ and $m_l$ are
the masses of the $B$ meson and the pseudoscalar meson, respectively.

Sometimes a momentum cutoff regularization scheme is also used to solve the
Bethe-Salpeter equation in the unitary coupled-channel
approximation. Thus the expression for $G_l$ is given
by~\cite{angels}
\begin{eqnarray}
G_{l} &=& i \, \int \frac{d^4 q}{(2 \pi)^4} \, \frac{1}{2E_l
(\vec{q})} \, \frac{1}{P^0 - q^0 - E_l (\vec{q}) + i \epsilon} \,
\frac{1}{q^2 - m^2_l + i \epsilon} \nonumber \\
&\rightarrow& \int_{|\vec q|<q_{max}} \, \frac{d^3 q}{(2 \pi)^3} \,
\frac{1}{2 \omega_l (\vec{q})} \, \frac{1}{2 E_l (\vec{q})} \,
\frac{1}{P^0 - \omega_l (\vec{q}) - E_l (\vec{q}) + i \epsilon}
\label{propcutoff}
\end{eqnarray}
with $P^0$ the total energy of the system, and $\omega_l (\vec{q})$,
$E_l (\vec{q})$ the energies of intermediate mesons and baryons,
respectively.

It is believed that the maximum momentum $q_{max}$ should be equal
to the regularization scale $\mu$ in the dimensional regularization
scheme approximately, and it has been proved that these two schemes always give
similar properties of the resonant state in the
meson-meson and meson-baryon interaction if we set $q_{max}\approx\mu$,
just as done in Refs.~\cite{jido, Molina, Geng, sun2010}.

In this work, the $B \bar{K}$ and $B_s \pi$ interaction will be studied by solving the Bethe-Salpeter equation using the loop function in Eq.~(\ref{eq:gpropdr}) and Eq.~(\ref{propcutoff}), respectively.

\section{Results}
\label{sect:result}

The leading order and next-to-leading order potentials between the $B$
meson and the pseudoscalar meson in different channels are depicted
in Fig.~\ref{fig:V_0629}. It is apparent that only in the channel of
$B\bar{K}\rightarrow B_s \pi$, the leading order potential is
attractive, while it is zero in the channels of $B \bar{K}
\rightarrow B \bar{K} $ and $B_s \pi \rightarrow B_s \pi$, as shown
in Table~\ref{table:C-LO}.
The next-to-leading order correction results in a repulsive potential
in the channels of $B \bar{K} \rightarrow B \bar{K} $ and $B_s \pi
\rightarrow B_s \pi$, and in the crossing channel
$B\bar{K}\rightarrow B_s \pi$, the next-to-leading order Lagrangian
gives a correction to the leading order potential.

If only the leading order potential of the $B$ meson and the
pseudoscalar meson in Eq.~(\ref{eq:LO}) is taken into account in the unitary
coupled-channel approximation, a
 pole of the squared amplitude $|T|^2$ appears at
$5567+i16$MeV in the complex energy plane with a momentum cutoff $q_{max}=3000$MeV according to the loop function in Eq.~(\ref{propcutoff}). Moreover, in the dimensional regularization scheme, a pole of the squared amplitude $|T|^2$ is generated dynamically at $5569+i16$MeV if we set the subtract constant $a=-2$ and the regularization scale $\mu=3100$MeV. It is apparent that the pole might be associated with the $X(5568)$ particle claimed by the D0 Collaboration. In addition, it is noticed that the momentum cutoff $q_{max}$ takes a similar value to the regularization scale $\mu$ if the subtract constant $a=-2$ is fixed.

If the next-to-leading order correction to the pseudoscalar meson and $B$ meson potential is taken into account,
a pole can be found at $5632+i25$MeV in the complex energy plane with $a=-2$ and $\nu=3100$MeV. Apparently, the real and imaginary parts of the pole are far away from the mass and decay width of the $X(5568)$ particle, respectively, so the subtract constant or the regularization scale has to be adjusted.
If we increase the value of the regularization scale $\mu$, while the subtract constant $a$ is fixed, the real and imaginary parts of the pole position both decrease.
A pole of the squared amplitude $|T|^2$ at $5570+i11$MeV is generated dynamically with $a=-2$ and $\mu=3700$MeV in the dimensional regularization scheme. However, in order to produce a pole in the same energy region with the loop function in Eq.~(\ref{propcutoff}), the momentum cutoff $q_{max}$ must be improved. Actually, with $q_{max}=4500$MeV, a pole of $|T|^2$ appears at $5568+i10$MeV in the complex energy plane.
Obviously, the value of $q_{max}$ in the momentum cutoff regularization scheme is far larger than the $\mu$ value in the dimensional regularization scheme if the next-to-leading order correction to the potential of the $B$ meson and the pseudoscalar meson is considered.

The pole position of the squared amplitude and the couplings of the
resonant state to $B\bar{K}$ and $B_s \pi$ channels are listed in
Table~\ref{table:coupling}. The label $LO+Dim$ denotes the results
calculated by using the leading order potential of the $B$ meson
and the pseudoscalar meson in the dimensional regularization scheme,
while the label $LO+Cutoff$ stands for the results calculated by using the leading order potential in the momentum cutoff regularization scheme.
The cases where the next-to-leading order correction is included in these two kinds of regularization schemes are labeled as $NLO+Dim$ and $NLO+Cutoff$, respectively.
From Table~\ref{table:coupling}, we can find that the pole position is above the $B_s \pi$ threshold and locates on the second Riemann sheet, which can be regarded as a resonant state of $B_s \pi$.
Moreover, the couplings of the
resonant state to $B \bar{K}$ and $B_s \pi$ channels in the
dimensional regularization scheme almost take the same values as those in the momentum cutoff regularization scheme, respectively.
Additionally, it is noticed that this resonant state couples more strongly to the $B
\bar{K}$ channel. Even if the next-to-leading order correction of the $B$ meson
and the pseudoscalar meson is taken into account, the coupling to the $B
\bar{K}$ channel is larger than the corresponding value to the $B_s \pi$ channel.

\begin{table}[hbt]
\begin{center}
\begin{tabular}{c|c|c|c|c}
\hline
                                   & LO+Dim            & LO+Cutoff           & NLO+Dim           & NLO+Cutoff          \\
\hline
    a                                & -2              & -              & -2             & -          \\
\hline
   $\mu$ or $q_{max}$ (MeV)          & 3100            & 3000           & 3700           & 4500  \\
\hline

 Pole position (MeV) & $5569+i16$     & $5567+i16$        & $5570+i11$    & $5568+i10$  \\
\hline
 $B \bar{K}$   & $138+i11$      & $140+i12$         & $112+i7$     & $111+i7$  \\
\hline
 $B_s \pi$     & $120+i19$      & $121+i20$        & $101+i13$     & $100+i13$   \\
\hline \hline
\end{tabular}
\end{center}
\caption{The pole position in the complex energy plane and the
couplings to different channel. The meaning of the labels $LO+Dim$, $LO+Cutoff$, $NLO+Dim$  and
$NLO+Cutoff$ can be found in the context.}
\label{table:coupling}
\end{table}

\section{$B^* \phi \rightarrow B^* \phi $}
\label{sect:Bstar}

From the Lagrangians in Eqs.~(\ref{eq:lolag}) and (\ref{eq:nlolag}), the leading order
and next-to-leading order potentials of the $B^*$ meson and the
pseudoscalar meson can be obtained similarly
\be V_{LO(NLO)}(P^*(p_1)+ \phi(k_1) \rightarrow P^*(p_2)+ \phi(k_2)
 ) = - \varepsilon \cdot \varepsilon^*
V_{LO(NLO)}(P(p_1)+ \phi(k_1) \rightarrow P(p_2)+ \phi(k_2),
 \ee
where $\varepsilon$ and $\varepsilon^*$ are the polarization vectors
of the initial and final $B^*$ mesons, respectively.
In the infinite heavy-quark limit,  $\varepsilon \cdot
\varepsilon^*=-1$\cite{Abreu:2011ic}. Thus the potential of the
$B^*$ meson with the pseudoscalar meson takes the same form as the
potential of the $B$ meson except that the mass of the $B$ meson is
replaced by the mass of the $B^*$ meson, correspondingly.

In the leading order approximation, we find a pole at $5620+i16$MeV
with isospin $I=1$ and spin $J=1$, and we have set $a=-2$ and
$\mu=3100$MeV in the dimensional regularization scheme.
If the next-to-leading order correction is taken into
account, moreover, the resonant state appears at $5620+i11$MeV with $a=-2$
and $\mu=3700$MeV. In the momentum cutoff regularization scheme, we obtain the pole position at $5616+i16$MeV with $q_{max}=3000$MeV if only the leading order potential of the $B^*$ meson and the pseudoscalar meson is taken into account. In addition, if the next-to-leading order correction is supplemented, a pole would appear at $5616+i10$MeV
with $q_{max}=4500$MeV.
Apparently, the pole is higher than the $B_s^* \pi$ threshold and lies on the second Riemann sheet.
If the claimed $X(5568)$ state with
$J^P=0^+$ is confirmed, there is also a resonant state with
$J^P=1^+$ and a mass around $5620MeV$.

\section{Partners in the charm sector}
\label{sect:charm}

Since the $D$ meson mass is less than the $B$ meson mass, the value of the regularization scale $\mu$ should be less than that of the $B$ meson-pseudoscalar meson interaction in the dimensional regularization scheme if the subtract constant $a=-2$ is fixed.
As to the charm sector of $DK$ and $D_s \pi$, we find a pole at
$2325+i65$MeV in the leading order approximation with $a=-2$ and
$\mu=1400$MeV, while a pole at $2326+i57$MeV in the complex energy
plane when we include the next-to-leading order correction in the $D$
meson-pseudoscalar meson interaction with
$a=-2$ and $\mu=1800$MeV.
In the momentum cutoff regularization scheme, the pole appears at $2349+i74$MeV with $q_{max}=1600$MeV if only the leading order potential is considered. Furthermore, if the next-to-leading order correction is taken into account, the pole lies at $2345+i62$MeV in the complex energy plane if the momentum cutoff $q_{max}=2500$MeV.
The pole position is above the $D_s \pi$ threshold and locates on the second Riemann sheet, thus it might correspond to a resonant state, which is much like the $D_s^*(2317)$ particle except that the isospin $I=1$.

For the $D^* \bar{K}$ and $D^*_s \pi$ sector, The pole is lower than the $D^* \bar{K}$ threshold and appears on the second Riemann sheet.
A pole of the squared amplitude $|T|^2$ is detected at $2493+i66$MeV in the complex energy plane if the leading order potential is taken into account in the dimensional regularization scheme, where we have set $a=-2$ and $\mu=1400$MeV. if the next-to-leading order correction is included, the pole moves to the position of $2490+i58$MeV with $a=-2$ and $\mu=1800$MeV.
In the momentum cutoff regularization scheme, the pole locates at $2502+i72$MeV with $q_{max}=1600$MeV in the leading order approximation.
When the next-to-leading order correction is taken into account, the pole appears at $2501+i61$MeV with the momentum cutoff $q_{max}=2500$MeV.

\section{Summary}
\label{sect:summary}

The possibility that the X(5568) announced by the D0 Collaboration
corresponds to a resonant state is discussed in this article.
The potential of the $B$ meson and the pseudoscalar meson is
deduced both in the leading order approximation and in the
leading order plus next-to-leading order approximation, and then
the amplitude of $B \bar{K}$ and $B_s \pi$ with isospin $I=1$ is
studied in the unitary coupled-channel approximation of
Bethe-Salpeter equation. By adjusting the value of the
regularization scale, we can obtain a reasonable pole of the
squared amplitude which can be associated with the $X(5568)$ state.

\section*{Acknowledgments}
We would like to thank Han-Qing Zheng, E. Oset and En Wang for
useful discussions.

\end{document}